\documentclass[onecolumn]{aastex62}

\usepackage{lineno}
\usepackage{longtable}

\graphicspath{./}

\begin{document}

\title{Observational properties of 155 O- and B-type massive pulsating stars}

\correspondingauthor{Xiang-dong Shi}
\email{sxd@ynao.ac.cn}

\author{Xiang-dong Shi}
\affiliation{Yunnan Observatories, Chinese Academy of Sciences(CAS), P.O. Box 110, 650216 Kunming, P.R. China}
\affiliation{University of Chinese Academy of Sciences, No.1 Yanqihu East Rd, Huairou District, Beijing, PR China, 101408}
\affiliation{Key Laboratory of the Structure and Evolution of Celestial Objects, CAS, Kunming 650216, P.R. China}

\author{Sheng-bang Qian}
\affiliation{Yunnan Observatories, Chinese Academy of Sciences(CAS), P.O. Box 110, 650216 Kunming, P.R. China}
\affiliation{University of Chinese Academy of Sciences, No.1 Yanqihu East Rd, Huairou District, Beijing, PR China, 101408}
\affiliation{Key Laboratory of the Structure and Evolution of Celestial Objects, CAS, Kunming 650216, P.R. China}

\author{Li-ying Zhu}
\affiliation{Yunnan Observatories, Chinese Academy of Sciences(CAS), P.O. Box 110, 650216 Kunming, P.R. China}
\affiliation{University of Chinese Academy of Sciences, No.1 Yanqihu East Rd, Huairou District, Beijing, PR China, 101408}
\affiliation{Key Laboratory of the Structure and Evolution of Celestial Objects, CAS, Kunming 650216, P.R. China}

\author{Liang Liu}
\affiliation{Yunnan Observatories, Chinese Academy of Sciences(CAS), P.O. Box 110, 650216 Kunming, P.R. China}
\affiliation{University of Chinese Academy of Sciences, No.1 Yanqihu East Rd, Huairou District, Beijing, PR China, 101408}
\affiliation{Key Laboratory of the Structure and Evolution of Celestial Objects, CAS, Kunming 650216, P.R. China}

\author{Lin-jia Li}
\affiliation{Yunnan Observatories, Chinese Academy of Sciences(CAS), P.O. Box 110, 650216 Kunming, P.R. China}
\affiliation{Key Laboratory of the Structure and Evolution of Celestial Objects, CAS, Kunming 650216, P.R. China}

\author{Lei Zang}
\affiliation{Yunnan Observatories, Chinese Academy of Sciences(CAS), P.O. Box 110, 650216 Kunming, P.R. China}
\affiliation{University of Chinese Academy of Sciences, No.1 Yanqihu East Rd, Huairou District, Beijing, PR China, 101408}
\affiliation{Key Laboratory of the Structure and Evolution of Celestial Objects, CAS, Kunming 650216, P.R. China}

\begin{abstract}
The O- and B-type (OB-type) pulsating stars are important objects to study the structure and evolution of massive stars through asteroseismology. A large amount of data from various sky surveys provide an unprecedented opportunity to search for and study this kind of variable star. We identify 155 OB-type pulsating stars or candidates, including 38 Oe/Be stars or candidates, from the data observed by TESS, LAMOST, and GAIA, which are almost new. Among the 155 objects, 87 samples are identified as SPB stars including 37 objects with pure low-frequency and 50 objects with both low- and high-frequency pulsation, and 14 samples are identified as BCEP stars with both low- and high-frequency pulsation. The H-R diagram shows that these SPB and BCEP stars are mainly located in their instability regions and in the evolutionary stage of the main-sequence with a mass range of 2.5-20 $M_{\odot}$ and 7-20 $M_{\odot}$. Two special objects show fourier spectra similar to BCEP stars but with different positions in H-R, Period-Temperature (P-T), and Period-Luminosity (P-L) diagrams. Meanwhile, 52 other targets are identified as candidates of OB-type pulsating stars. We also derive the preliminary results of the P-L relation for SPB and BCEP stars, respectively. This work also indicates that in addition to the H-R diagram, P-T and P-L diagrams are also very useful for the classification of SPB and BCEP. Further detailed analysis of these objects can dramatically increase our understanding of theories of evolution and structure for massive OB-type pulsating stars.
\end{abstract}


\section{Introduction} \label{sec:intro}

Massive stars generally refer to O- and B-type (OB-type) stars, which have the characteristics of high mass, temperature, and luminosity \citep{1968ApJ...151..611M, 1973ARA&A..11...29M, 1973AJ.....78..929P}. They are related to many important objects and physical processes, such as neutron stars, black holes, supernovae, and gravitational wave events \citep{2010ApJ...725..940Y, 2008ApJ...676.1162S, 2020RAA....20..161H, 2020A&A...638A..39L}. Despite recent advances, there are still many deficiencies in the study of massive stars, such as internal element-mixing, stellar wind mass loss, magnetic activity, internal angular momentum transport, and so on. To make up for these deficiencies, it is very important to study the structure and evolution of massive stars by the method of asteroseismology, because it is the only way to obtain their internal structure information at present.

Slowly pulsating B stars (SPB stars) and $\beta$ Cephei pulsating variable stars (BCEP stars) are the only two recognized classes of OB-type pulsating variable stars in the upper main-sequence. The SPB stars are generally with a spectral type of late B-type (about B3 to B9) and a mass from 2.5 to 8 $M_{\odot}$. They pulsate in non-radial multiperiodic g-mode with a period range of 0.5 to 3 days \citep{2010aste.book.....A}. BCEP stars oscillate in low-order p- and g-mode \citep{2005ApJS..158..193S} with a typical period range of 2 to 7 hours. They are late O-type or early B-type stars (earlier than B5) with a roughly mass range of 7 to 20 $M_{\odot}$ \citep{1993MNRAS.265..588D}. The pulsations of SPB stars and BCEP stars seem to be excited by the $\kappa$ mechanism driven by the ionization zone of iron-group elements.

\clearpage

\begin{longtable}{llllllllllll}

\caption{\label{tab:1} The Catalog of OB-type massive pulsating stars observed by TESS, LAMOST, and GAIA.}\\
\hline
TESS ID     & $\pi$    & V       & $Teff$  & Flag   & $log(L_/L_{\odot})$ & Fourier  & Comments & Period     & Amplitude \\
            & ($mas$)  & ($Mag$) & ($K$)   &        &                     & spectra  &          & ($days$)   & ($mmag$)  \\

\hline
\endfirsthead

\caption{\label{tab:frequency}(Continued)}\\
\hline
TESS ID     & $\pi$    & V       & $Teff$  & Flag   & $log(L_/L_{\odot})$ & Fourier & Comments & Period     & Amplitude \\
            & ($mas$)  & ($Mag$) & ($K$)   &        &                     & spectra &          & ($days$)   & ($mmag$)  \\

\hline \endhead

\hline
\multicolumn{6}{r}{\textsl{(Continued)}}\\
\endfoot

\hline

\endlastfoot

2944381   & 0.57 & 9.91   & 12080(179)     & 1      & 2.79  & LF       & SPB      & 1.6182(2)    & 2.52(1)  \\
2945216   & 0.30 & 10.70  & 27206(811)     & 1      & 3.94  & LF       & SPB      & 4.316(2)     & 0.88(1)  \\
3030849   & 0.52 & 10.29  & 14399(198)     & 1      & 3.00  & LF       & SPB      & 1.34811(9)   & 6.34(3)  \\
4136259   & 2.94 & 7.76   & 12891(272)     & 1      & 2.20  & LF       & SPB      & 0.84957(5)   & 1.61(1)  \\
5612096   & 6.37 & 5.36   & 16256          & 2      & 2.73  & LF       & SPB(C)   & 1.4735(1)    & 14.48(8) \\
45646950  & 0.52 & 12.79  & 16622(321)     & 1(Be)  & 3.16  & LF       & SPB      & 0.7793(1)    & 3.44(4)  \\
45709015  & 0.68 & 9.95   & 18428(419)     & 1      & 3.32  & LF       & SPB      & 1.8932(6)    & 1.13(2)  \\
53327951  & 0.47 & 8.72   & 15546          & 2(Be)  & 3.75  & LF       & SPB      & 1.0217(2)    & 3.32(3)  \\
55079633  & 0.29 & 9.89   & 25885(888)     & 1      & 4.13  & LF       & SPB      & 9.145(7)     & 2.72(2)  \\
63748979  & 0.73 & 12.24  & 13590          & 2(Be)  & 2.51  & LF       & SPB(C)   & 0.8179(2)    & 8.3(1)   \\
68989297  & 2.13 & 8.70   & 17314(338)     & 1      & 2.43  & LF       & SPB      & 2.28948(7)   & 17.74(1) \\
75508636  & 0.95 & 9.62   & 17944(194)     & 1      & 2.93  & LF       & SPB      & 1.3141(2)    & 1.59(2)  \\
77881231  & 0.78 & 10.94  & 11412(111)     & 1      & 2.05  & LF       & SPB      & 1.7827(1)    & 15.28(4) \\
96951421  & 0.60 & 9.22   & 14500          & 4(Be)  & 3.22  & LF       & SPB      & 1.13355(7)   & 28.8(1)  \\
127584684 & 0.48 & 10.32  & 20788(485)     & 1      & 3.38  & LF       & SPB      & 4.346(3)     & 0.75(1)  \\
137815897 & 0.45 & 10.15  & 18486(281)     & 1      & 3.28  & LF       & SPB(C)   & 1.09276(8)   & 1.72(1)  \\
138366301 & 0.53 & 10.00  & 15909(232)     & 1      & 3.04  & LF       & SPB      & 2.558(3)     & 0.67(2)  \\
155305056 & 0.21 & 9.14   & 21035          & 2(Be?) & 4.55  & LF       & SPB      & 0.62176(8)   & 5.09(4)  \\
158662338 & 0.34 & 12.96  & 11705(131)     & 1      & 1.94  & LF       & SPB(C)   & 1.3604(4)    & 19.7(2)  \\
172426403 & 1.09 & 9.34   & 16207(277)     & 1      & 2.71  & LF       & SPB      & 1.5994(5)    & 2.52(2)  \\
190059323 &  1.38 & 8.15  & 9549           & 2      & 2.45  & LF       & SPB      & 1.352(1)     & 0.15(1)           \\
202270396 & 0.42 & 12.75  & 13030          & 2      & 2.04  & LF       & SPB(C)   & 0.6205(1)    & 8.2(1)   \\
236101031 & 0.55 & 9.57   & 20258(329)     & 1      & 3.46  & LF       & SPB      & 2.927(1)     & 4.67(3)  \\
247638066 & 3.64 & 5.99   & 12210          & 2      & 2.68  & LF       & SPB      & 3.8000(1)    & 1.43(1)  \\
252287651 & 0.93 & 8.86   & 13956(1237)    & 1      & 2.94  & LF       & SPB      & 3.218(2)     & 0.94(1)  \\
261361813 & 2.04 & 8.33   & 13168(177)     & 1      & 2.34  & LF       & SPB      & 3.982(2)     & 1.45(1)  \\
269079069 & 2.12 & 7.98   & 13906(188)     & 1      & 2.48  & LF       & SPB      & 1.114714(4)  & 14.38(6) \\
273537365 & 1.36 & 9.82   & 15470(264)     & 1      & 2.36  & LF       & SPB      & 1.6774(7)    & 2.54(3)  \\
286043331 & 1.15 & 9.93   & 13528(214)     & 1      & 2.26  & LF       & SPB      & 3.0586(5)    & 1.95(1)  \\
293929714 & 3.49 & 6.39   & 16849          & 2      & 2.91  & LF       & SPB      & 1.5409(3)    & 5.89(4)  \\
320273520 & 0.97 & 9.52   & 13502(163)     & 1      & 2.55  & LF       & SPB      & 2.59(1)      & 0.13(1)  \\
328231445 & 7.34 & 5.61   & 13040          & 2      & 2.28  & LF       & SPB      & 0.33485(4)   & 0.19(1)  \\
355826643 & 1.64 & 8.80   & 16335          & 2      & 2.66  & LF       & SPB      & 1.6335(2)    & 0.88(1)  \\
404745241 & 3.00 & 7.68   & 12913(245)     & 1      & 2.23  & LF       & SPB      & 1.2076(2)    & 5.36(4)  \\
426419494 & 0.52 & 13.02  & 12830(184)     & 1      & 1.63  & LF       & SPB      & 0.48409(1)   & 7.76(4)  \\
429750051 & 0.55 & 9.06   & 24332          & 2      & 4.18  & LF       & SPB      & 3.991(2)     & 6.16(6)  \\
712312017 & 3.15 & 8.25   & 12321(202)     & 1      & 1.92  & LF       & SPB      & 0.35934(6)   & 1.74(4)  \\
1827744   & 1.01 & 9.82   & 14999(202)     & 1      & 2.52  & LF+HF    & SPB      & 0.433897(4)  & 2.85(1)  \\
2334539   & 0.95 & 9.37   & 20005(407)     & 1      & 3.06  & LF+HF    & SPB      & 1.7416(6)    & 0.49(1)  \\
3097389   & 0.49 & 9.26   & 20000(51)      & 1      & 3.79  & LF+HF    & SPB      & 3.668(3)     & 0.32(1)  \\
17126940  & 0.53 & 13.49  & 32447          & 2      & 3.28  & LF+HF    & SPB      & 3.340(4)     & 1.28(2)  \\
19935717  & 0.76 & 10.77  & 15590(247)     & 1(Be)  & 2.56  & LF+HF    & SPB      & 0.48104(3)   & 2.32(3)  \\
20309687  & 0.78 & 10.19  & 16893(269)     & 1      & 2.87  & LF+HF    & SPB      & 0.43838(3)   & 1.08(1)  \\
23992299  & 0.11 & 13.21  & 15398(206)     & 1      & 3.03  & LF+HF    & SPB      & 1.1823(5)    & 5.13(8)  \\
27847920  & 0.34 & 10.62  & 16785(295)     & 1      & 3.24  & LF+HF    & SPB      & 0.483028(2)  & 1.99(2)  \\
56750848  & 1.16 & 8.64   & 22741(4422)    & 1(Be)  & 3.24  & LF+HF    & SPB      & 5.175(3)     & 4.97(6)  \\
59468515  & 1.06 & 10.07  & 17883(383)     & 1      & 2.60  & LF+HF    & SPB      & 0.47572(2)   & 1.11(1)  \\
69225473  & 3.39 & 7.96   & 14504          & 2      & 2.14  & LF+HF    & SPB      & 2.4936(3)    & 2.79(1)  \\
76543999  & 0.99 & 8.22   & 18519          & 2(Be?) & 3.43  & LF+HF    & SPB      & 0.47182(1)   & 7.30(4)  \\
76641076  & 0.27 & 9.72   & 25029(518)     & 1      & 4.47  & LF+HF    & SPB      & 0.57454(4)   & 5.74(6)  \\
78140949  & 1.00 & 9.73   & 14047(237)     & 1      & 2.57  & LF+HF    & SPB      & 0.38759(2)   & 1.72(1)  \\
78261043  & 0.40 & 9.45   & 19379          & 2      & 3.87  & LF+HF    & SPB      & 3.949(2)     & 2.90(4)  \\
80980016  & 0.49 & 9.39   & 21190(426)     & 1      & 3.78  & LF+HF    & SPB      & 3.716(2)     & 0.97(1)  \\
81259541  & 0.97 & 8.46   & 23542          & 2      & 3.52  & LF+HF    & SPB      & 0.344228(7)  & 13.35(7) \\
117103922 & 0.83 & 8.95   & 20748(341)     & 1      & 3.41  & LF+HF    & SPB      & 0.82294(8)   & 1.33(1)  \\
162616331 & 0.37 & 11.66  & 14676(179)     & 1      & 2.58  & LF+HF    & SPB      & 0.45571(5)   & 0.51(2)  \\
200516204 & 1.40 & 9.48   & 12490(175)     & 1(Be)  & 2.15  & LF+HF    & SPB      & 0.45750(2)   & 4.85(3)  \\
234704735 & 0.38 & 11.30  & 14587(476)     & 1(Be)  & 3.05  & LF+HF    & SPB      & 0.33594(3)   & 4.32(5)  \\
234853418 & 1.10 & 9.82   & 14904(293)     & 1      & 2.52  & LF+HF    & SPB      & 0.438725(1)  & 1.44(1)  \\
234923805 & 0.60 & 12.68  & 19494          & 2(Be)  & 2.34  & LF+HF    & SPB      & 2.573(2)     & 2.71(4)  \\
234927025 & 1.43 & 9.20   & 13937(367)     & 1(Be)  & 2.70  & LF+HF    & SPB      & 0.5878(1)    & 0.88(1)  \\
234933597 & 0.70 & 7.74   & 15050          & 2(Be)  & 4.46  & LF+HF    & SPB      & 1.65737(2)   & 2.53(2)  \\
250137613 & 2.92 & 6.42   & 15174          & 2      & 2.91  & LF+HF    & SPB      & 1.327400(4)  & 6.80(2)  \\
264613665 & 0.41 & 11.20  & 22876(462)     & 1      & 3.22  & LF+HF    & SPB      & 2.466(2)     & 2.61(3)  \\
266656195 & 1.14 & 7.77   & 18049(1749)    & 1(Be)  & 3.37  & LF+HF    & SPB      & 3.015(3)     & 10.6(2)  \\
269267602 & 0.57 & 10.86  & 17514(306)     & 1      & 2.80  & LF+HF    & SPB      & 1.85131(2)   & 2.60(2)  \\
269617027 & 0.94 & 9.36   & 13700(262)     & 1      & 2.76  & LF+HF    & SPB      & 1.09744(6)   & 5.24(4)  \\
273038264 & 0.27 & 10.96  & 22179(353)     & 1      & 3.62  & LF+HF    & SPB      & 0.7574(4)    & 0.83(3)  \\
282207882 & 0.71 & 9.43   & 15014(370)     & 1      & 3.08  & LF+HF    & SPB      & 0.33785(3)   & 2.66(3)  \\
309335864 & 0.78 & 9.16   & 19477(566)     & 1(Be)  & 3.28  & LF+HF    & SPB      & 0.54676(4)   & 2.00(2)  \\
358301744 & 1.08 & 8.98   & 15907(193)     & 1      & 2.89  & LF+HF    & SPB      & 0.49892(4)   & 0.53(1)  \\
427352300 & 2.79 & 7.60   & 12767          & 2      & 2.33  & LF+HF    & SPB      & 5.69(1)      & 0.45(1)  \\
9745058   & 1.10 & 10.72  & 18856          & 3(Be)  & 2.36  & HF+LF    & SPB      & 0.26122(1)   & 2.94(3)  \\
72520779  & 0.49 & 11.62  & 13574(179)     & 1      & 2.42  & HF+LF    & SPB      & 0.33241(1)   & 12.40(5) \\
78499882  & 0.60 & 10.00  & 15407          & 2(Be)  & 2.98  & HF+LF    & SPB      & 0.28280(1)   & 5.25(6)  \\
117007503 & 0.78 & 10.15  & 18234(332)     & 1      & 2.89  & HF+LF    & SPB      & 0.204284(3)  & 2.81(2)  \\
129011169 & 1.79 & 9.89   & 11013(113)     & 1      & 1.64  & HF+LF    & SPB      & 0.27051(2)   & 1.59(2)  \\
201960635 & 0.49 & 11.02  & 17633(442)     & 1(Be)  & 2.91  & HF+LF    & SPB      & 0.2364358(4) & 1.92(2)  \\
234172192 & 0.70 & 9.72   & 14367(474)     & 1      & 2.93  & HF+LF    & SPB      & 0.330332(8)  & 12.81(4) \\
408991617 & 0.94 & 10.14  & 15981(256)     & 1      & 2.52  & HF+LF    & SPB      & 0.173038(9)  & 1.66(2)  \\
414916124 & 1.37 & 10.93  & 11896(135)     & 1      & 1.60  & HF+LF    & SPB      & 0.32183(2)   & 4.09(3)  \\
446041643 & 1.89 & 9.52   & 15266(196)     & 1      & 2.10  & HF+LF    & SPB      & 0.222960(2)  & 6.81(2)  \\
20412106  & 1.12 & 8.25   & 13419(190)     & 1      & 2.97  & LF+HF    & SPB(C)   & 0.50573(2)   & 0.58(1)  \\
68816670  & 2.07 & 8.61   & 15567(199)     & 1      & 2.38  & LF+HF    & SPB      & 0.93473(6)   & 2.41(2)  \\
115730751 & 0.83 & 10.39  & 16942(416)     & 1      & 2.66  & LF+HF    & SPB      & 2.9687(7)    & 1.37(2)  \\
8300380   & 1.47 & 9.45   & 16642          & 2      & 2.50  & HF+LF    & SPB      & 0.199347(8)  & 0.63(3)  \\
76211045  & 0.60 & 9.61   & 21521(367)     & 1      & 3.61  & LF+HF    & SPB      & 0.31331(3)   & 0.40(1)  \\
29277573  & 0.81 & 8.60   & 14858          & 2(Be)  & 3.24  & LF+HF    & BCEP     & 0.0882321(6) & 2.41(2)  \\
45802921  & 0.56 & 10.37  & 22296(644)     & 1      & 3.46  & LF+HF    & BCEP     & 0.107432(4)  & 0.35(1)  \\
48217508  & 0.46 & 9.84   & 18387(388)     & 1      & 3.36  & LF+HF    & BCEP     & 0.0957064(9) & 1.29(1)  \\
12672166  & 0.55 & 9.78   & 24374(1144)    & 1      & 3.79  & HF+LF    & BCEP     & 0.145089(5)  & 1.32(2)  \\
45799839  & 0.54 & 10.16  & 21409(722)     & 1      & 3.55  & HF+LF    & BCEP     & 0.132007(1)  & 4.05(3)  \\
232846315 & 0.47 & 9.46   & 20000(46)      & 1      & 3.68  & HF+LF    & BCEP     & 0.1318001(2) & 4.28(5)  \\
234834992 & 0.67 & 8.16   & 24102          & 3      & 4.25  & HF+LF    & BCEP     & 0.1653974(4) & 0.33(1)  \\
234879286 & 0.62 & 11.14  & 23537(684)     & 1      & 2.98  & HF+LF    & BCEP     & 0.076697(3)  & 1.01(2)  \\
272625912 & 0.35 & 10.00  & 25740(526)     & 1      & 4.26  & HF+LF    & BCEP     & 0.3025623(6) & 15.4(1)  \\
336987555 & 0.95 & 9.14   & 21740(517)     & 1      & 3.43  & HF+LF    & BCEP     & 0.123298(3)  & 0.35(1)  \\
348506791 & 0.39 & 8.50   & 30419(637)     & 1      & 4.63  & HF       & BCEP(C)  & 0.217349(6)  & 13.9(1)  \\
348608238 & 0.43 & 10.62  & 24449(452)     & 1      & 3.53  & HF+LF    & BCEP(C)  & 0.176728(8)  & 2.30(2)  \\
348671609 & 0.39 & 10.01  & 21794(331)     & 1      & 3.74  & HF+LF    & BCEP(C)  & 0.226561(7)  & 5.82(3)  \\
372115570 & 0.42 & 10.09  & 24377(399)     & 1      & 3.80  & HF+LF    & BCEP(C)  & 0.252373(9)  & 5.25(3)  \\
9153664   & 1.92 & 8.73   & 12548          & 2      & 2.16  & HF       & BCEP/SPB & 0.217741(3)  & 16.11(4) \\
458654855 & 6.68 & 5.91   & 12228          & 2      & 2.18  & HF+LF    & BCEP/SPB & 0.157130(1)  & 0.50(1)  \\

\end{longtable}
{Note. The numbers in the parentheses are the errors of the data. These targets with the $H_{\alpha}$ emission line profile detected from LAMOST are marked as Be in the parentheses of Column 5. These capital letters C in the parentheses of Column 8 represent these targets that may be contaminated by neighboring stars in the TESS photometry apertures.\\}


\citet{2005ApJS..158..193S} published a catalog of 93 confirmed BCEP stars including 14 stars discovered by \citet{2005AcA....55..219P}. Three new BCEP stars were detected by \citet{2019MNRAS.489.1304B} from the $K2$ mission. Recently, the KELT exoplanet survey detected a total of 113 BCEP stars including 86 new discoveries \citep{2020AJ....160...32L}. Some researchers have made ground-based surveys for SPB stars, such as \citet{1999A&A...343..872A, 2001A&A...379..905M, 2002A&A...393..965D, 2007A&A...463..243D}. More than 15 new SPB candidates were identified from the CoRoT space mission \citep{2009A&A...506..471D}. \citet{2011MNRAS.413.2403B} detected 15 pulsating stars showing the low-frequency characteristics of SPB stars from Kepler space mission \citep{2010Sci...327..977B}, of which 7 stars also showed some weak and isolated high-frequency.

NASA launched the Transiting Exoplanet Survey Satellite (TESS, \citet{2015JATIS...1a4003R}) in April 2018. The all-sky is divided into 26 sectors, including 13 sectors of the southern sky and northern sky, respectively. Each sector is observed continuously for 27 days, and the all-sky can be repeated for about two years. Every sector with the 24*96 degree field-of-view provides light curves with a 2-min short cadence for about 20,000 targets and also provides full-frame images every 30 minutes. The primary TESS mission is to detect Earth-sized planets transiting bright and nearby stars in all-sky. At the same time, its massive and high-precision photometric data also provides an unprecedented opportunity for the research of binaries and variable stars, especially for bright variables (e.g., \citet{2020MNRAS.493.5871B, 2022ApJS..259..50S}). Most recently, \citet{2020MNRAS.493.5871B} published a catalog of 766 pulsating main-sequence B-type stars observed by TESS in sectors 1-18, which greatly expands the existing samples.

The Large Sky Area Multi-Object Fiber Spectroscopic Telescope (LAMOST, also called as Guo Shou Jing telescope) is a special quasi-meridian reflecting Schmidt telescope located at the Xinglong Station of National Astronomical Observatory in China \citep{1996ApOpt..35.5155W, 2012RAA....12.1197C}. It has a field of view of 5 square degrees and could obtain 4000 spectra in a single exposure. The telescope is equipped with a low-resolution spectrograph and a medium-resolution spectrograph. The low-resolution spectrograph has a wavelength range of 3700-9000 $\AA$, including both a blue arm (3700-5900 $\AA$) and a red arm (5700-9000 $\AA$). The medium-resolution spectrograph is also divided into both a blue arm (4950-5350 $\AA$) and a red arm (6300-6800 $\AA$). LAMOST has obtained a large number of stellar spectral data (e.g., \citet{2015RAA....15.1095L, 2012RAA....12..723Z}), including many binary systems (e.g., \citet{2017RAA....17...87Q, 2018ApJS..235....5Q, 2020RAA....20..163Q, 2019ApJS..244...43Z}), many pulsating stars (e.g., \citet{2018MNRAS.475..478Q, 2019RAA....19....1Q}) and more than two hundred and fifty thousand candidates of binaries or variable stars \citep{2019RAA....19...64Q}.

Gaia Satellite was launched in December 2013 and is a cornerstone mission of the European Space Agency (ESA). It has provided very high-precision astrometry data for nearly 2 billion stars \citep{2016A&A...595A...1G, 2018A&A...616A...1G, 2021A&A...649A...1G}. Its parallax data is a key parameter to calculate the luminosity of stars, which provides an important way to research variables (e.g., \citet{2019MNRAS.485.2380M}) or to test the reliability of research results (e.g., \citet{2021AJ....161...46S, 2021PASP..133e4201S}).
Gaia DR3 also derived the stellar atmospheric parameters from the BP/RP and RVS spectrum by different methods, such as the General Stellar Parametrizer from Photometry (GSP-Phot) and the Extended Stellar Parametrizer for Hot Stars (ESP-HS).

In this paper, we identify 155 OB-type pulsating stars observed by TESS, LAMOST, and GAIA, and their observational and physical properties are also analyzed. Firstly, the TESS light curves of these samples are analyzed by Fourier spectrum. Secondly, some statistical analyses are made according to their parallax information and effective temperatures.

\begin{figure*}\centering \vbox to4.0in{\rule{0pt}{5.0in}}
\includegraphics{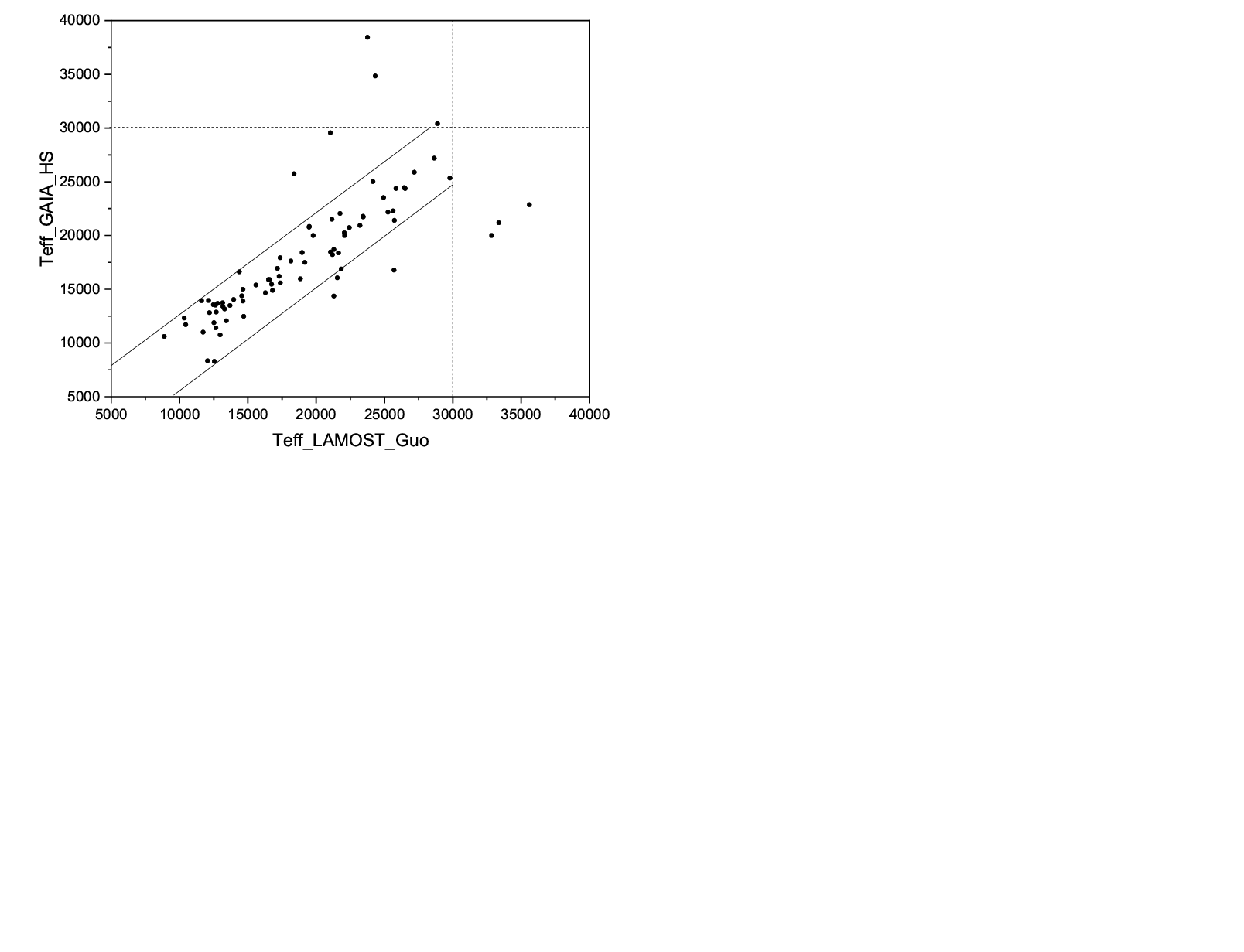}
\caption{The comparison between the temperature of Gaia ESP-HS and that of LAMOST derived by \citet{2021ApJS..257...54G} (LAMOST$\_$Guo). Most of these targets give consistent results.}
\label{fig:T-T}
\end{figure*}

\begin{figure*}\centering \vbox to4.0in{\rule{0pt}{5.0in}}
\includegraphics{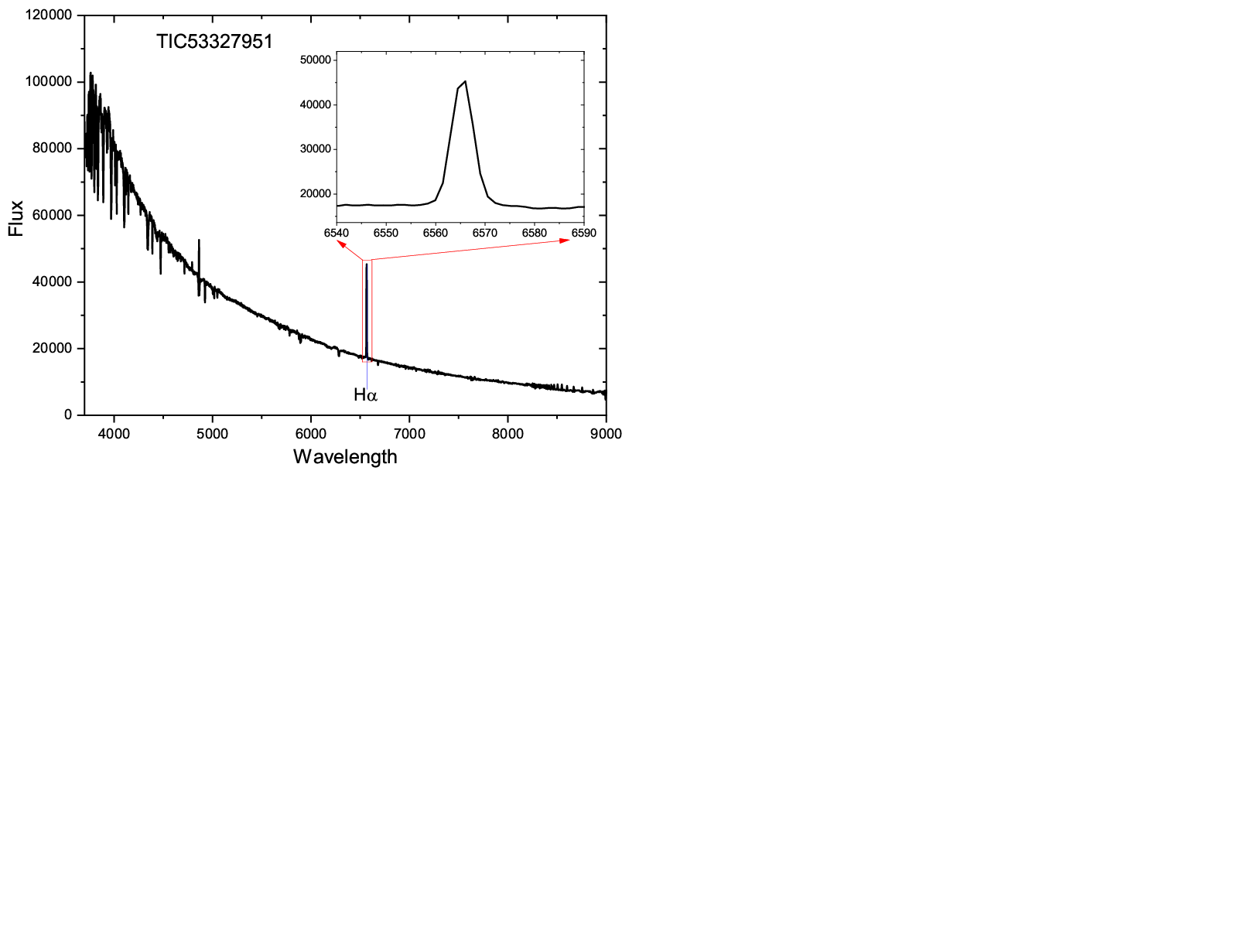}
\caption{Example low-resolution spectra of Be star TIC53327951 from LAMOST.}
\label{fig:Spectra}
\end{figure*}

\begin{figure*}\centering \vbox to9.5in{\rule{0pt}{5.0in}}
\includegraphics{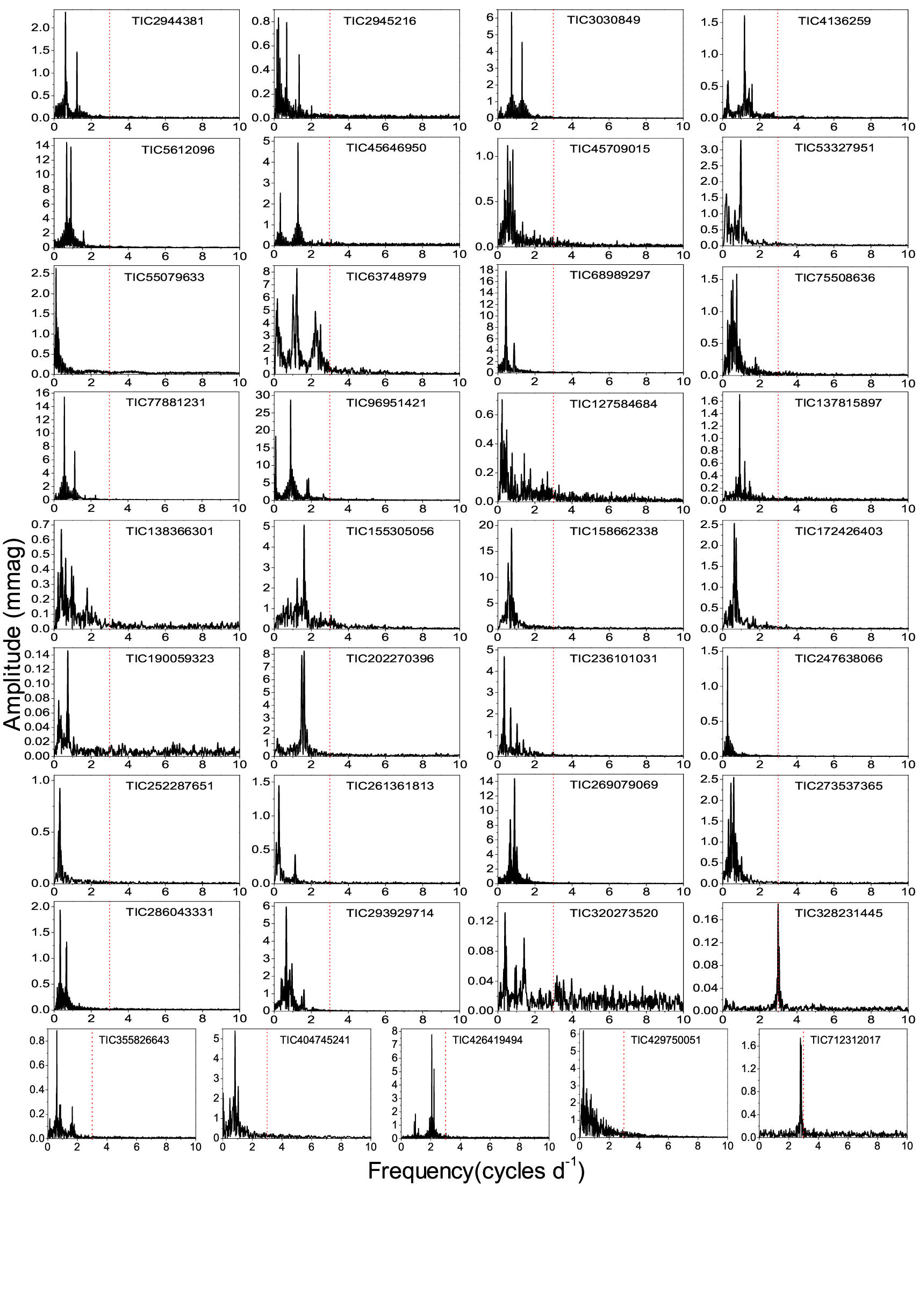}
\caption{Fourier spectra of the OB-type pulsating stars with pure low-frequency (LF).}
\label{fig:1}
\end{figure*}


\begin{figure*}\centering \vbox to9.0in{\rule{0pt}{5.0in}}
\includegraphics{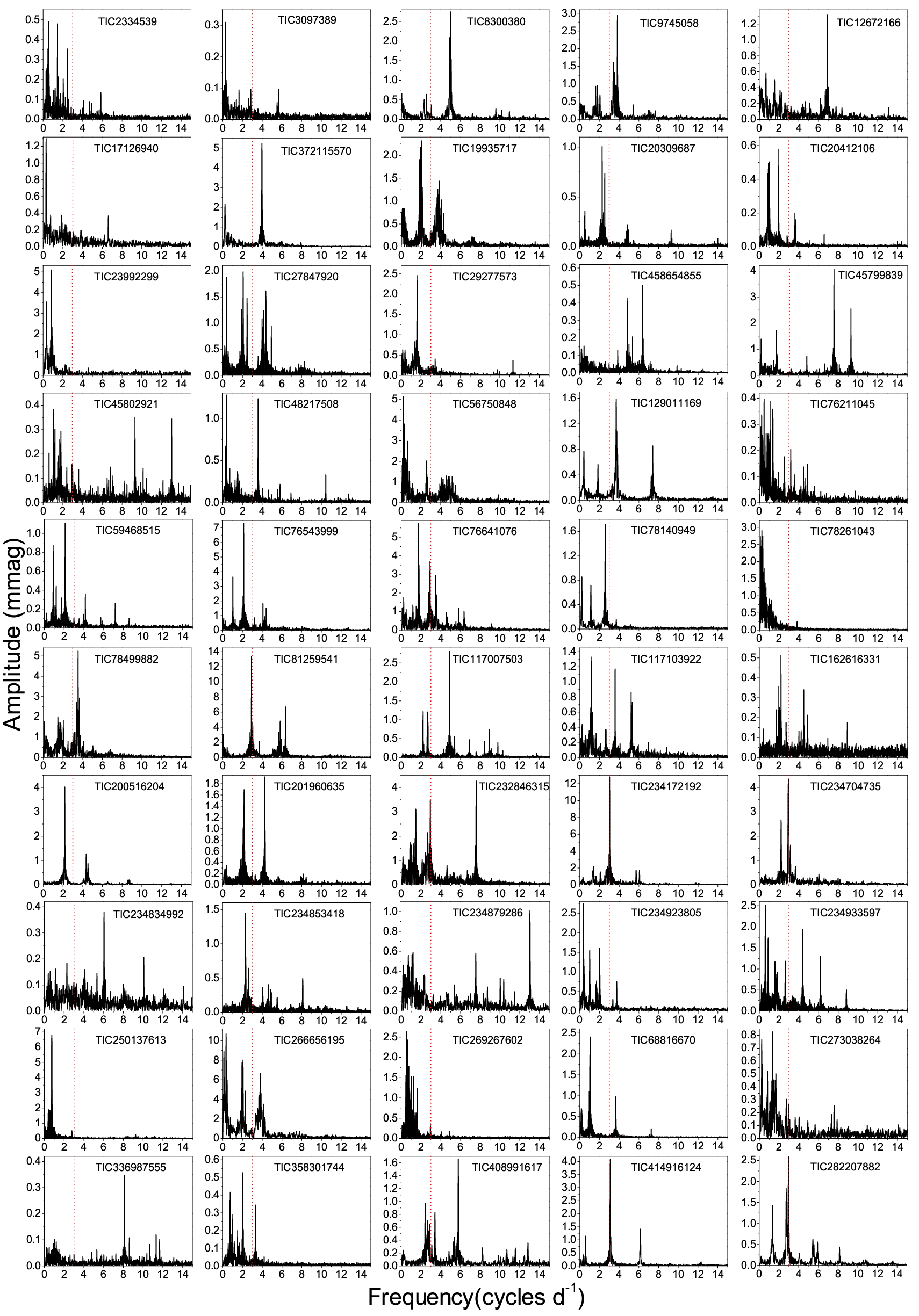}
\caption{Fourier spectra of the OB-type pulsating stars with high-frequency (HF), part 1.}
\label{fig:2}
\end{figure*}

\begin{figure*}\centering \vbox to4.0in{\rule{0pt}{5.0in}}
\includegraphics{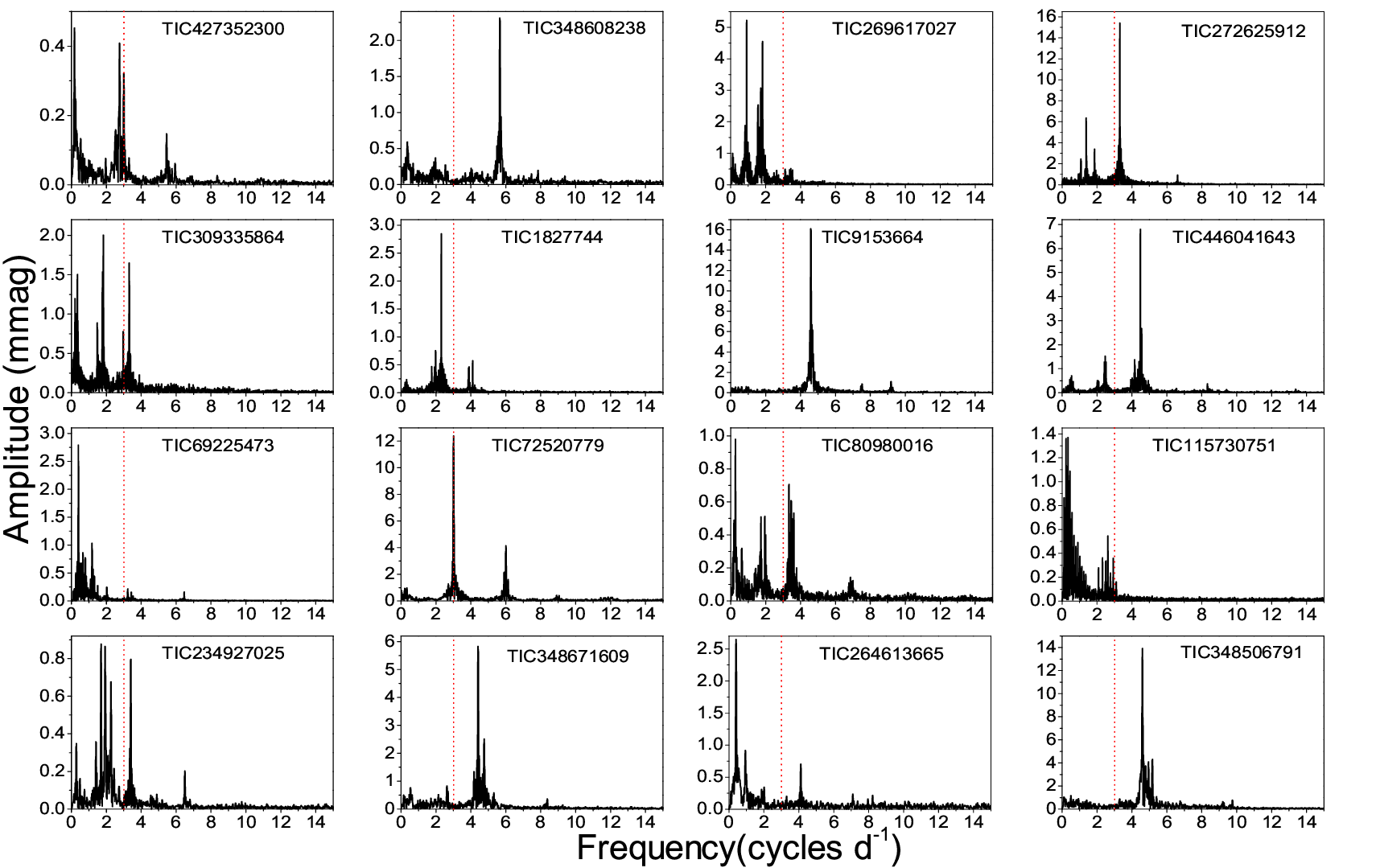}
\caption{Fourier spectra of the OB-type pulsating stars with high-frequency (HF), part 2.}
\label{fig:3}
\end{figure*}

\section{The OB-type pulsating stars observed by TESS, LAMOST, and GAIA} \label{}

From the Mikulski Space Telescope Archive (MAST) database, We downloaded the light curves observed by TESS with a 2-min short cadence. Then, we choose to use the light curves that have been corrected for long-term drifts by using pre-search data conditioning (PDC, \citet{2010ApJ...713L..87J}) and process them using the steps described by \citet{2021AJ....161...46S, 2021MNRAS.505.6166S}. A total of 155 OB-type pulsating stars or candidates are identified by a simple program aided visual classification. The information is listed in Table \ref{tab:1} and \ref{tab:2} for these pulsating stars or candidates, respectively. We have also carried on cross-match these samples with the catalog of 766 pulsating main-sequence B-type stars \citep{2020MNRAS.493.5871B}, and a total of 10 objects are listed in this catalog and shown in Table \ref{tab:3}.

Meanwhile, the parallax $\pi$ of these samples are observed by Gaia Survey and listed in Columns 2 of Table \ref{tab:1} and \ref{tab:2}.
The visual magnitude $m_{V}$ of these samples are from the TESS Input Catalogue (TIC, \citet{2018AJ....156..102S}) or SIMBAD \citep{2000A&AS..143....9W} and are listed in Columns 3 of Table \ref{tab:1} and \ref{tab:2}. Whether there are neighboring stars contaminating the TESS photometry apertures is checked on SIMBAD, and mark the targets that may be contaminated with a capital letter C in the parentheses at Column 8 of Table \ref{tab:1} and Column 7 of \ref{tab:2}.

Gaia ESP-HS and GSP-Phot derived the surface effective temperatures for most of these samples. Meanwhile, LAMOST observed the low- or medium-resolution spectra of most of these samples, and their stellar atmospheric parameters were given by \citet{2021ApJS..257...54G}. Fig \ref{fig:T-T} shows the comparison between the temperature of Gaia ESP-HS and that of LAMOST derived by \citet{2021ApJS..257...54G} (LAMOST$\_$Guo), where these targets with temperatures below 30000 $K$ give consistent results. This means that their temperature should be reliable except for individual targets. However, the consistency of other atmospheric parameters is very poor, so only the effective temperature is quoted in this paper.

The effective temperatures $Teff$ of these samples are listed in Column 4 of Table \ref{tab:1} and \ref{tab:2}, and Column 5 assigns a priority flag according to the method of deriving $Teff$. First, second, third, and fourth priority is given to Gaia ESP-HS, LAMOST$\_$Guo, Gaia GSP-Phot, and the temperature estimation from the spectral type given by LAMOST, respectively. The uncertainties of effective temperature given by Gaia ESP-HS are listed in the parentheses of Columns 4 in Table \ref{tab:1} and \ref{tab:2}. The average uncertainties of effective temperature estimated by \citet{2021ApJS..257...54G} are $\sigma_{Teff} = 1642 K$  for low-resolution spectra and $\sigma_{Teff} = 2185 K$ for medium-resolution spectra, respectively. The effective temperatures from low-resolution spectra have smaller uncertainty than that from medium-resolution spectra, which may be due to the wider wavelength coverage of the low-resolution spectra \citep{2021ApJS..257...54G}. Those surface effective temperatures given by third and fourth priority may lead to more errors than those given by \citet{2021ApJS..257...54G}.

We check the $H_{\alpha}$ emission line profile from LAMOST data and marked these Oe/Be stars and candidates (with a question mark) in the parentheses of Column 5 of Table \ref{tab:1} and \ref{tab:2}. Fig \ref{fig:Spectra} shows an example of low-resolution spectra from LAMOST for a Be star. The existence of the emission lines may affect the effective temperature determinations.


\begin{table*}[]
\begin{center}
\caption{The candidates of OB-type pulsating stars observed by TESS, LAMOST, and GAIA.}
\label{tab:2}
\begin{tabular}{lllllllllllllll}
 \hline
TESS ID     & $\pi$    & V       & $Teff$  & Flag   & $log(L_/L_{\odot})$  & Comments \\
            & ($mas$)  & ($Mag$) & ($K$)   &        &                      &          \\

\hline

16700873  &  0.14 & 13.00  & 16607(502)    & 1      & 3.06   & ?           \\
16732768  & 0.58  & 10.64  & 29346         & 2      & 3.65   & ?           \\
17125739  &  0.54 & 11.12  & 26061         & 2      & 3.59   & ?(C)        \\
17449204  &  0.52 & 10.96  & 12333         & 2(Be?) & 3.51   & ?           \\
17449469  &  0.55 & 12.40  & 32610         & 2(Oe)  & 3.93   & ?           \\
45717172  &  0.57 & 7.43   & 22590         & 2      & 4.58   & ?           \\
62706478  &  1.38 & 8.20   & 13660(367)    & 1      & 2.74   & ?           \\
63366439  &  0.60 & 10.77  & 13160(163)    & 1      & 2.41   & SPB+ROT     \\
63457231  &  0.79 & 9.88   & 24689         & 2(Be?) & 3.74   & ?           \\
75509134  &  1.19 & 8.92   & 22063(281)    & 1(Be)  & 3.20   & ?           \\
76216928  &  0.80 & 9.51   & 16065(288)    & 1(Be)  & 3.03   & ?           \\
78431470  &  0.63 & 7.04   & 22653         & 2      & 4.53   & ?           \\
78515820  &  0.23 & 10.89  & 20182         & 2      & 4.11   & ?           \\
121067499 &  0.94 & 8.12   & 20315         & 2      & 3.61   & ?           \\
121605728 &  2.35 & 4.40   & 15997         & 2      & 3.97   & ?           \\
127586563 &  0.31 & 10.90  & 25722(631)    & 1      & 4.12   & SPB+ROT     \\
139424697 &  0.45 & 12.03  & 25090         & 2      & 2.77   & ?           \\
143216039 &  1.41 & 8.11   & 10400         & 2      & 2.55   & ?           \\
143860583 &  1.21 & 8.63   & 17400(357)    & 1      & 3.01   & ?           \\
167031605 &  0.99 & 9.91   & 10919(158)    & 1      & 2.22   & SPB+EB      \\
200519617 &  1.22 & 7.51   & 21303(1856)   & 1(Be)  & 3.60   & ?           \\
216236227 &  0.35 & 8.52   & 23329         & 2      & 4.62   & ?(C)        \\
234038269 &  1.32 & 9.93   & 11695(198)    & 1      & 2.02   & ?           \\
234841110 &  1.39 & 9.27   & 25345(846)    & 1(Be?) & 3.05   & ?(C)        \\
234933368 &  0.47 & 12.46  & 19341         & 2(Be)  & 3.31   & ?           \\
237668110 &  1.39 & 7.85   & 19127(1283)   & 1(Be)  & 3.26   & SPB+ROT     \\
239880242 &  2.36 & 8.23   & 11867(267)    & 1(Be)  & 2.15   & ?           \\
245048006 &  1.08 & 9.25   & 13911(162)    & 1      & 2.76   & ?           \\
261534342 &  1.09 & 8.73   & 14500         & 4      & 2.88   & ?           \\
272368236 &  0.57 & 7.50   & 29111         & 2      & 4.57   & SPB+EB      \\
272843619 &  0.28 & 10.61  & 18703(304)    & 1      & 3.54   & ?(C)        \\
290646079 &  0.27 & 11.40  & 13741(129)    & 1      & 2.90   & ?           \\
302538146 &  3.40 & 7.00   & 14725         & 2(Be)  & 2.55   & SPB+ROT     \\
317144497 &  0.41 & 9.33   & 12288         & 2      & 3.80   & ?           \\
319574477 &  0.63 & 9.16   & 26558(964)    & 1      & 3.82   & ?           \\
319655114 &  0.47 & 9.01   & 36131(1596)   & 1      & 4.39   & ?(C)        \\
320935848 &  0.14 & 11.39  & 20950(10)     & 1(Be)  & 3.95   & ?           \\
334155902 &  1.81 & 9.72   & 10619(102)    & 1      & 1.67   & ?           \\
343878395 &  0.65 & 10.14  & 10700         & 4(Be)  & 4.22   & SPB+ROT     \\
344241254 &  5.35 & 6.66   & 9760          & 2      & 1.88   & ?           \\
348023238 &  0.44 & 9.18   & 27335(534)    & 1      & 4.14   & BCEP+ROT(C) \\
349070699 &  0.67 & 8.36   & 23474         & 2(Be)  & 3.89   & SPB+ROT     \\
366657610 &  6.68 & 7.56   & 11392         & 2      & 1.45   & ROT+SPB     \\
412187395 &  1.84 & 9.20   & 12712(144)    & 1(Be)  & 2.04   & ?           \\
415674825 &  1.47 & 10.62  & 12056         & 2      & 1.60   & ?           \\
429715043 &  0.52 & 10.72  & 23159         & 2(Be)  & 3.40   & SPB+ROT     \\
429749401 &  0.51 & 8.79   & 32098(458)    & 1(Oe)  & 4.65   & SPB+ROT     \\
434303361 &  0.46 & 9.79   & 12302         & 2      & 3.31   & ?(C)        \\
436722939 &  0.63 & 7.80   & 12926         & 2      & 3.60   & SPB+ROT     \\
445988959 &  2.73 & 7.05   & 14017         & 2(Be)  & 2.64   & ?           \\
470314334 &  0.99 & 10.01  & 14856(240)    & 1      & 2.54   & SPB+EB(C)   \\
470472700 &  0.20 & 9.69   & 23936         & 2(Be?) & 4.98   & ?           \\

\hline
\end{tabular}
\end{center}
\tablecomments{Note. The numbers in the parentheses are the errors of the data. These targets with the $H_{\alpha}$ emission line profile detected from LAMOST are marked as Be/Oe in the parentheses of Column 5. These capital letters C in the parentheses of Column 7 represent these targets that may be contaminated by neighboring stars in the TESS photometry apertures.\\}
\end{table*}

\begin{table*}[]
\begin{center}
\caption{The OB-type pulsating stars and candidates in the catalog of \citet{2020MNRAS.493.5871B}.}
\label{tab:3}
\begin{tabular}{lllllllllllllll}
 \hline
TESS ID     & RA         & DEC      \\
            &($deg$)     & ($deg$)  \\
\hline
48217508  & 283.11978 & 48.40394    \\
232846315 & 311.55274 & 35.54043    \\
234853418 & 98.06834  & 1.17470     \\
234933597 & 98.47670  & 4.65978     \\
245048006 & 302.47686 & 27.89957    \\
250137613 & 63.40896  & -1.14962    \\
269267602 & 299.79242 & 44.86245    \\
282207882 & 102.95928 & -5.61977    \\
328231445 & 45.59591  & 4.35269     \\
427352300 & 83.63496  & -4.48949    \\
\hline
\end{tabular}
\end{center}
\end{table*}

\section{The classification of these OB-type pulsating stars}

The fourier spectra of these OB-type pulsating samples are analyzed using the Period04 software \citep{2005CoAst.146...53L} that are based on classical Fourier analysis. The signal-to-noise ratio (S/N) of frequency is calculated with a box size of 1 $d^{-1}$ and the residuals at original, and the errors of frequencies and amplitudes are calculated according to \citet{1999DSSN...13...28M}.

We assume the value 3 cycles $d^{-1}$ as a rough boundary between low-frequency (LF) and high-frequency (HF) in the fourier spectra, according to the properties of p mode and g mode (e.g., \citet{2010aste.book.....A, 2005ApJS..158..193S}), and the classification of these variable stars will consider more later. Columns 8 of Table \ref{tab:1} give the descriptions of the fourier spectra and the recommended variable star type, respectively. A star is classified as SPB if it only has the LF, i.e., there is not a frequency of more than 3 cycles $d^{-1}$ with an S/N greater than 5.4 for a 2-min short cadence \citep{2021AcA....71..113B}.

In many cases, a star is detected to have LF and HF at the same time, which will be marked as LF+HF or HF+LF according to which range frequencies seem to dominate. If there are several independent high-frequencies, a star should be a BCEP star, because BCEP stars can oscillate in p- and g-mode \citep{2005ApJS..158..193S} or those low-frequencies are just combination frequencies of high-frequencies. On the contrary, it is classified as SPB, if those high-frequencies are combination frequencies of low-frequencies. For example, \citet{2015MNRAS.450.3015K} attribute frequency groups in the fourier spectra to g-mode pulsations and combination frequencies. There are two targets (TIC9153664 and TIC458654855) that should be classified as BCEP according to their fourier spectra but are marked as BCEP/SPB, which will be described in the next section.

These samples identified as SPB, BCEP, and BCEP/SPB are listed in Table \ref{tab:1}, and their fourier spectra are shown in Fig \ref{fig:1}-\ref{fig:3}. Their dominant frequencies and amplitudes are listed in Columns 9-10 of Table \ref{tab:1}, and the errors on the last one bits are in parentheses. We found that almost all targets with HF also have LF, except TIC9153664 and TIC348506791. Moreover, it seems that the two targets also have some frequencies less than 3 cycles $d^{-1}$ from the fourier spectra, but their S/N is less than 5.4. This may be because less data coverage (only one sector) of TIC9153664 and TIC348506791 makes these low-frequencies with a low S/N.

Generally, rotational variability is caused by spots or the stellar wind \citep{2020A&A...639A..81B}, etc, which in fourier spectra shows as a series (possibly more than two) of harmonic frequencies of the rotation frequency, similar to the eclipsing binary, but no obvious eclipse can be seen in the light curve. Table \ref{tab:2} lists 52 candidate targets, including the following situations:

a) a pulsating variable star may be in an eclipsing binary (EB) or a rotational variable (ROT), including SPB+EB and BCEP+EB or SPB+ROT and BCEP+ROT,

b) if a variable star could not detect coherent periodic signals and is suspected of exhibiting internal gravity waves or rotational variation, it will be marked as a question mark.

\section{The properties of these OB-type pulsating stars}

\begin{figure*}\centering \vbox to4.0in{\rule{0pt}{5.0in}}
\includegraphics{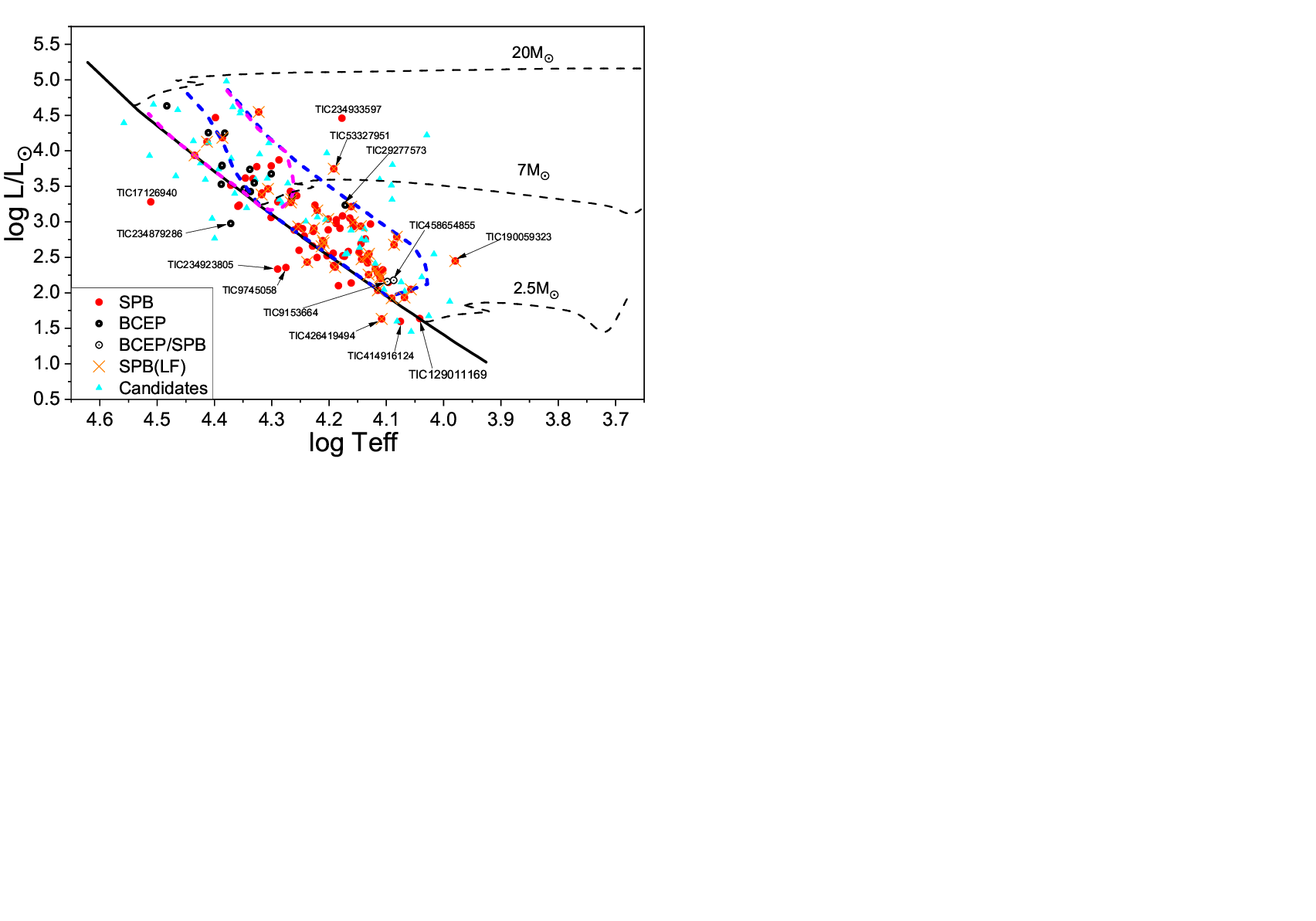}
\caption{The H-R diagram of these OB-type pulsating stars. The red solid circles, the black open circles, and the black open circles with dots refer to these SPB, BCEP, and BCEP/SPB stars, respectively. The orange crosses represent those SPB stars with pure low-frequency. Meanwhile, the candidate targets are shown as cyan solid triangles. The black solid and dotted lines are the theoretical zero-age main sequence and the evolutionary tracks for the different masses with Z=0.02. The blue and magenta dotted line is the instability regions of the SPB and BCEP stars for Z = 0.02 and spherical harmonic degree $l \leq$ 3 from \citet{2007CoAst.151...48M}, respectively.}
\label{fig:L-T}
\end{figure*}

\begin{figure*}\centering \vbox to4.0in{\rule{0pt}{5.0in}}
\includegraphics{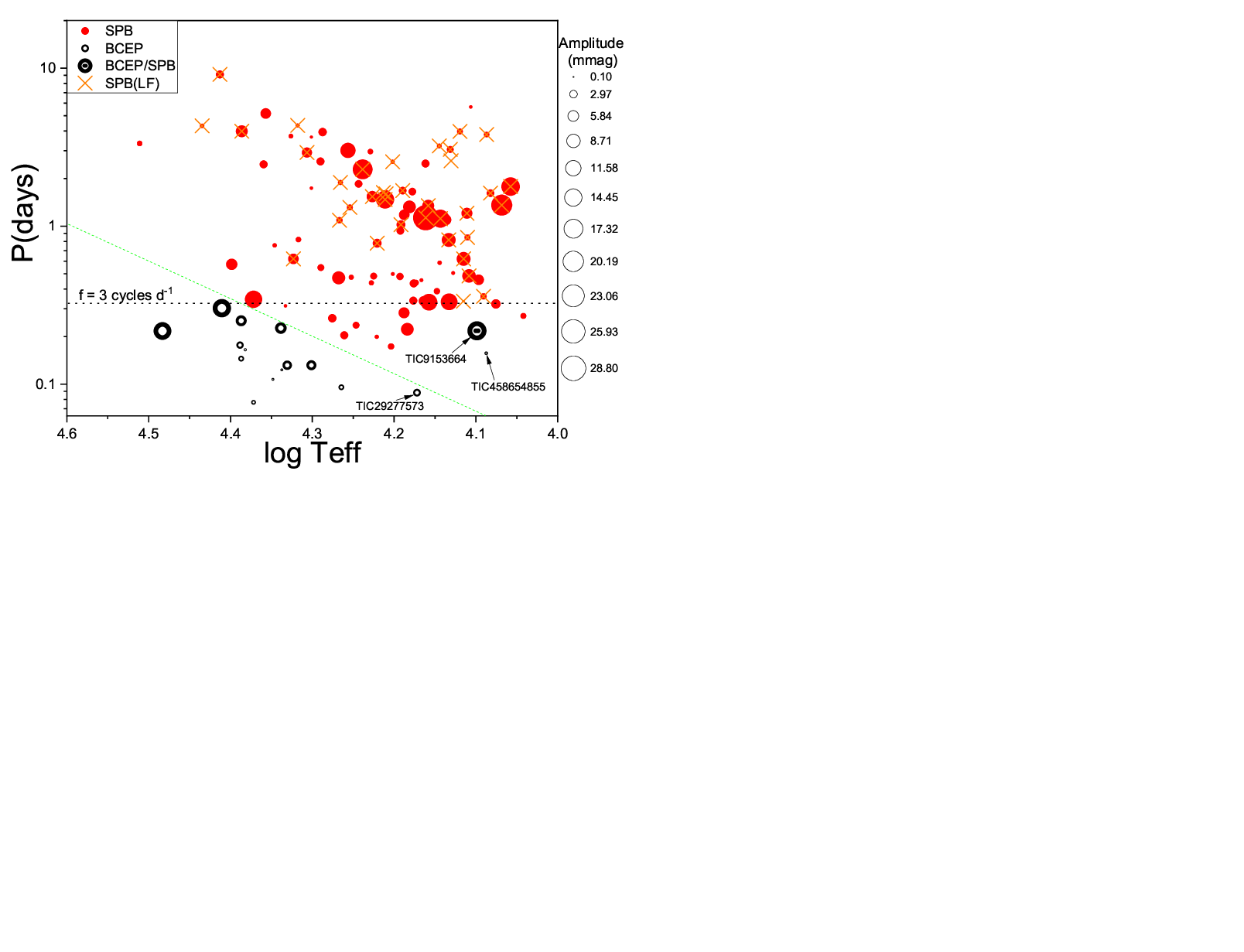}
\caption{The dominant pulsating period and the effective temperature relation diagram of these OB-type pulsating stars. The horizontal dashed lines represent a frequency = 3 cycles $d^{-1}$. Symbols are similar to those in Fig \ref{fig:L-T}, but the size of the circles indicates their pulsation amplitude of the dominant frequency.}
\label{fig:P-T}
\end{figure*}

\begin{figure*}\centering \vbox to4.0in{\rule{0pt}{5.0in}}
\includegraphics{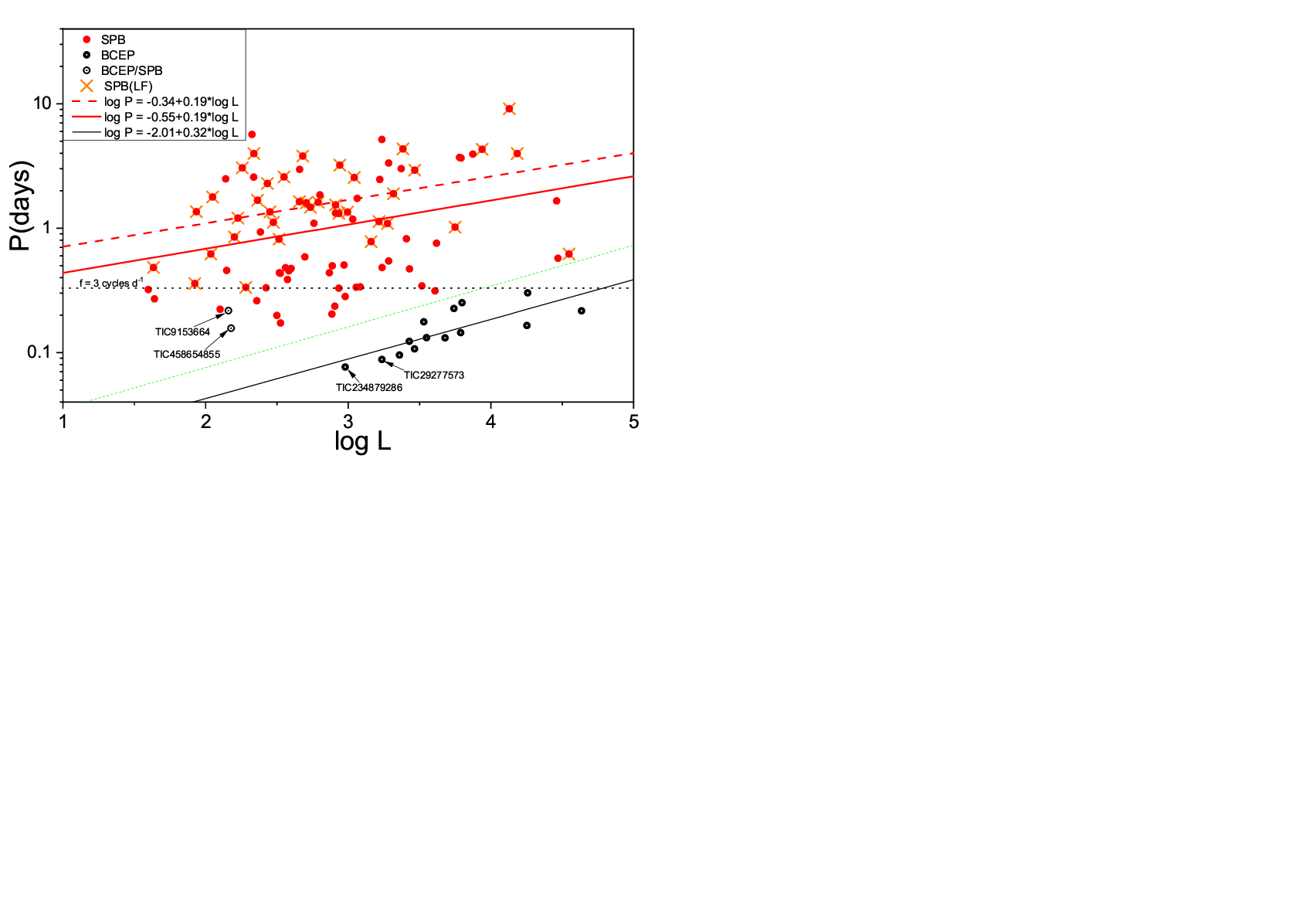}
\caption{The P-L diagram of these OB-type pulsating stars. The horizontal dashed lines represent a frequency = 3 cycles $d^{-1}$. Symbols are the same as those in Fig \ref{fig:L-T}. The red solid line, the red dash line, and the black solid line refer to the relation derived by the least-square method for all SPB stars, those SPB stars with pure low-frequency, and those BCEP stars.}
\label{fig:P-L}
\end{figure*}

The luminosity of these OB-type pulsating stars or candidates are shown in Columns 6 of Table \ref{tab:1} and \ref{tab:2}, which are calculated using
\begin{equation}
log(L_/L_{\odot})=0.4\times{(4.74-M_{V}-BC)}
\end{equation}
\begin{equation}
M_{V}=m_{V}-5\times{log(1000/\pi)}+5-A_{V},
\end{equation}
where the bolometric correction $BC$ is estimated based on the surface effective temperature using the calibration derived by \citet{2013ApJS..208....9P}, and the interstellar extinction $A_{V}$ is calculated from \citet{2011ApJ...737..103S}. The average uncertainties of $log(L_/L_{\odot})$ is estimated as about 0.1 dex, from the standard deviations for the visual magnitude $m_{V}$ about 0.01 mag, for the bolometric correction $BC$ about 0.02 mag, for the interstellar extinction $A_{V}$ about 0.10 mag, and for the Gaia parallax about 0.05 mas.

The H-R diagram of these OB-type pulsating stars and candidates is displayed in Fig \ref{fig:L-T}. In this figure, the evolutionary tracks for the mass of 2.5, 7, and 20 $M_{\odot}$, and the theoretical zero-age main sequence (ZAMS) for Z = 0.02 were created with the stellar evolution code Modules for Experiments in Stellar Astrophysics (MESA, \citet{2011ApJS..192....3P, 2013ApJS..208....4P, 2015ApJS..220...15P, 2018ApJS..234...34P, 2019ApJS..243...10P}. The blue and magenta dotted line is the instability regions of the SPB and BCEP stars for Z = 0.02 and spherical harmonic degree $l \leq$ 3 from \citet{2007CoAst.151...48M}, respectively.

We rechecked some targets that deviated from the instability regions in the H-R diagram. It is found that the effective temperature of some targets may be unreliable, such as TIC234933597, 53327951, 29277573, 190059323, 9745058, and 17126940 do not have the effective temperature given by Gaia ESP-HS, only by LAMOST$\_$Guo or Gaia GSP-Phot with low reliability. However, the effective temperature for most targets should be reliable, even for some targets that deviate significantly in the H-R diagram. For example, Gaia ESP-HS and LAMOST$\_$Guo both gave the same effective temperature for TIC234879286, 426419494, 414916124, 234923805, and 129011169, and their effective temperature should be reliable.

Fig \ref{fig:P-T} shows the relation between the dominant pulsation period and the surface effective temperature (P-T) for these OB-type pulsating stars, where the size of the circles indicates their pulsation amplitude of the dominant frequency. Except TIC9153664 and TIC458654855, although there are a few targets with an unreliable effective temperature, those SPB and BCEP stars can be separated by a simple straight line in the P-T diagram.


The relation of their dominant pulsation period and luminosity (P-L) is shown in Fig \ref{fig:P-L}, in which those SPB and BCEP stars show rough linear distributions. The least-square method derived the fitting equation, for all SPB stars:
\begin{equation}\label{eq:P-L1}
log P = -0.55(\pm0.20)+0.19(\pm0.07)\times{log L},
\end{equation}
for those SPB stars with pure low-frequency:
\begin{equation}\label{eq:P-L2}
log P = -0.34(\pm0.21)+0.19(\pm0.07)\times{log L},
\end{equation}
and for those BCEP stars:
\begin{equation}\label{eq:P-L3}
log P = -2.01(\pm0.27)+0.32(\pm0.07)\times{log L}.
\end{equation}
Equations \ref{eq:P-L1} and \ref{eq:P-L2} have the same slope but are different from equation \ref{eq:P-L3}, indicating that those SPB with pure low-frequency and with combination frequencies in high-frequency belong to the same kind of variables and are different from those of BCEP stars. Similar to the P-T diagram, those SPB and BCEP stars can also be separated by a simple straight line in the P-L diagram.

As can be seen in the H-R diagram, these SPB and BCEP stars are mainly located in their instability regions, and most are in the evolutionary stage of the main sequence with a mass range of 2.5-20 $M_{\odot}$ for SPB and 7-20 $M_{\odot}$ for BCEP.

TIC9153664 and 458654855 both show fourier spectra similar to those BCEP stars, but they are not only distributed in the same region as those SPB stars in the H-R diagram but also in the P-T and P-L diagrams. Therefore, they are marked as BCEP/SPB.


\section{Discussion and conclusion}

We identify 155 OB-type pulsating stars or candidates, including 38 Oe/Be stars or candidates, by visual classification from the data of TESS, LAMOST, and GAIA. Then, we calculated their luminosity, and obtain their H-R diagram by combining it with their effective temperature. Among the 155 objects, 87 samples are identified as SPB stars including 37 objects with pure low-frequency and 50 objects with both low- and high-frequency pulsation, where these high-frequency pulsations are combined by low-frequencies. A total of 14 targets with both low-frequency and high-frequency are identified as BCEP stars, and our results also indicate that there is no pure high-frequency pulsating star. The H-R diagram shows that these SPB and BCEP stars are mainly located in their instability regions and the evolutionary stage of the main sequence with a mass range of 2.5-20 $M_{\odot}$ for SPB and 7-20 $M_{\odot}$ for BCEP.

Except for those targets with unreliable temperatures, there are a few stars located around the instability regions in the H-R diagram, which may be the pulsating subdwarfs or the main sequence pulsating dwarfs with a different metallicity. Some objects (such as TIC23992299, 16700873, 290646079, and 426419494) are located 20-50 degrees off the galactic plane, indicating they should not be young objects, possibly because they are massive stars originating from binary mergers (e.g., \citet{2016MNRAS.457.2355S} and \citet{2022NatAs...6..480W}) or runaway stars from the galactic plane. Of course, they could also simply be subdwarfs as classified in SIMBAD.

TIC9153664 and 458654855 both show fourier spectra similar to that of BCEP stars, but they are not only distributed in the same region as those SPB stars in the H-R diagram but also in the P-T and P-L diagrams. If their positions in the H-R diagram are correct, they are likely to be a special kind of target. Their high-frequencies may be likely g-mode frequencies shifted to higher apparent values by rapid rotation (e.g., \citet{2013MNRAS.429.2500B}).In addition, the results of Stromgren-Crawford uvby$\beta$ photometry \citep{2015A&A...580A..23P} seem to indicate that both stars are of spectral type A and rather evolved, which is consistent with their relatively high luminosity as main-sequence A-type stars and their locations outside the Galactic plane (+19 and +28 degrees, respectively). In this case, they could just as well simply be $\delta$ Sct pulsators.

We also derive the preliminary results of the P-L relation for SPB and BCEP stars by the least-square method, which may have the potential for the massive pulsating stars as a standard candle to measure distances. However, more samples and further research are needed to improve their accuracy, especially for BCEP stars.

Although the SPB and BCEP stars have overlapping regions in the H-R diagram, they can be distinguished in the P-T and P-L diagrams. In addition to the H-R diagram, P-T and P-L diagrams are also very useful for the classification of SPB and BCEP.

These OB-type pulsating stars are very important sources for further study of massive stars from the method of asteroseismology, which has the potential to dramatically increase our understanding of the theory of evolution and structure for this type of star. In the future, it is necessary to study these objects from various views in more detail.

\acknowledgments
This work is partly supported by Chinese Natural Science Foundation (Nos. 11933008, 12103084 and 12273103) and the basic research project of Yunnan Province (Grant No. 202201AT070092).
The spectral data of this paper were observed by the Large Sky Area Multi-Object Fiber Spectroscopic Telescope (LAMOST).
This work has made use of data from the European Space Agency (ESA) mission Gaia (https://www.cosmos.esa.int/gaia), processed by the Gaia Data Processing and Analysis Consortium (DPAC, https://www.cosmos.esa.int/web/gaia/dpac/consortium). Funding for the DPAC has been provided by national institutions, in particular the institutions
participating in the Gaia Multilateral Agreement.
The TESS data presented in this paper were obtained from the Mikulski Archive for Space Telescopes (MAST) at the Space Telescope Science Institute (STScI). STScI is operated
by the Association of Universities for Research in Astronomy, Inc. Support to MAST for these data is provided by the NASA Office of Space Science. Funding for the TESS mission is provided by the NASA Explorer Program.
We thank the anonymous reviewer for his valuable comments and suggestions, which have greatly improved our manuscript.


\end{document}